\documentclass{aa}
\usepackage{psfig}
\usepackage{epsfig}
\def \sax {BeppoSAX}

\def \nh {N${\rm _H}$}

\def \hcm {\hbox {\ifmmode $ atom cm$^{-2}\else atom cm$^{-2}$\fi}}
\def \arcmin {\hbox{$^\prime$}}

\def \chisq {$\chi ^{2}$}
\def \rchisq {$\chi_{\nu} ^{2}$}
\def\approxgt{\mathrel{\hbox{\rlap{\lower.55ex \hbox {$\sim$}}
        \kern-.3em \raise.4ex \hbox{$>$}}}}
\def\approxlt{\mathrel{\hbox{\rlap{\lower.55ex \hbox {$\sim$}}
        \kern-.3em \raise.4ex \hbox{$<$}}}}

\begin{document}

\thesaurus{06.(02.01.2; 08.09.2; 08.14.1; 13.25.1; 13.25.3)}

\title{Simultaneous BeppoSAX and RXTE observations of the X-ray
burst sources GX 3+1 and  Ser X-1}

\author{T. Oosterbroek\inst{1}
        \and D. Barret\inst{2} 
        \and M. Guainazzi\inst{3} 
		\and E.C. Ford\inst{4} 
}
\offprints{T. Oosterbroek (toosterb@astro.estec.esa.nl)}

\institute{Astrophysics Division, Space Science Department of ESA, ESTEC,
       Postbus 299, NL-2200 AG Noordwijk, The Netherlands
\and
Centre d' Etude Spatiale des Rayonnements, CNRS/UPS, 9 Avenue du
       Colonel Roche, 31028 Toulouse Cedex 04, France
\and
XMM-Newton SOC, VILSPA, ESA, Apartado 50727, E-28080 Madrid, Spain
\and
Astronomical Institute, ``Anton Pannekoek,''
       University of Amsterdam, Kruislaan 403, 1098 SJ Amsterdam, The
       Netherlands.
}
\date{Received 25 July 2000 / Accepted 30 October 2000 }

\titlerunning{BeppoSAX and RXTE observations of GX 3+1 and Ser X-1}
\maketitle

\begin{abstract}

We have obtained spectral and timing data on GX 3+1 and Ser X-1. Both
sources were observed simultaneously with \sax\ and RXTE. The RXTE
data is used to provide power spectra and colour-colour diagrams in
order to constrain the state (and thus track $\dot M$) the sources are
in. The \sax\ data provide the broad-band spectra. The spectra of both
sources are reasonably well-fit using a model consisting of a
disk-blackbody, a comptonized component and a Fe line, absorbed by
interstellar absorption. The electron temperature (kT$_{\rm e}$) of
the Comptonizing plasma is in both cases $\sim$2.5 keV. This implies
that no strong high-energy tail from the Comptonized component
is present in either of the sources. We discuss the similarities
between these burst sources and the luminous X-ray sources located in
globular clusters.  We find that the spectral parameters of the
comptonized component provide information about the mass-accretion
rate, which agrees well with estimates from the timing and spectral
variations.


\end{abstract}

\section{Introduction}
\label{sect:intro}
The sources GX 3+1 and Ser X-1 are both bright X-ray burst sources.
Ser X-1 was discovered in 1965 (Friedmann et al.\ 1967), and
X-ray bursts from this source were first detected by Swank et al.\
(1976). GX 3+1 was discovered by Bradt et al.\ (1968), and the
first X-ray bursts were detected by Makishima et al.\ (\cite{m:83}).

These sources, containing neutron stars, are studied with \sax\ in
order to quantify their spectrum in the $\sim$0.1--30 keV energy
range. It is now well established that hard X-ray emission (above
$\sim$30 keV) is not an unique feature of black hole systems as
indicated by detections of several X-ray binaries at high ($\sim$100
keV) energies (e.g., Barret \& Vedrenne 1994). Following the
detections of neutron star systems at high energies van Paradijs \&
van der Klis (\cite{vpvdk:94}) concluded from data collected from the
HEAO 1 A-4 catalogue that the 10--80 keV spectra of accreting neutron
stars with low magnetic fields get progressively harder when the
source gets dimmer.  Historically observations with Ginga and nearly
simultaneous with BATSE of 4U 1608-522 (Yoshida et al. \cite{y:93},
Zhang et al. \cite{z:96}) have shown hard spectra from a neutron star
when it is in a low state.  A recent example of a neutron star showing
a high energy-tail in its spectrum is 4U 0614+091 (Piraino et al.\
\cite{psfk:99})

However, quantative
differences might exist between the two types of compact objects. 
For example Churazov et al.\ (\cite{c:95}) suggest that the power-law
components of black-hole candidates are systematically harder than
those of low-luminosity neutron star accretors.

The
two sources are observed in a program aimed at determining and
quantifying the spectral properties in the 0.1--30 keV band. Results
on data obtained with RXTE on four different X-ray bursters have been
reported by Barret et al.\ (2000). They find that the spectra of
sources in a low state ($\sim$0.05--0.1 L$_{\rm Edd}$) can be
described by thermal Comptonization with an electron temperature of
25--30 keV and (in general) an underlying soft component. When the
sources are in a higher state ($\sim$0.35 L$_{\rm Edd}$) there is no
significant hard tail, while the Comptonized component can be
described by a much lower electron temperature, kT$_{\rm e}$ ($\sim$3
keV).

We use (quasi-)simultaneous RXTE data to constrain the state of the
sources. According to Hasinger \& van der Klis (\cite{hk:89})
``Atoll'' sources trace out a distinct pattern in a colour-colour
diagram, while their power spectra show distinct changes when moving
through these states. This will allow us to obtain an estimate of the
immediate mass-accretion rate of these sources. Using the simultaneous
data obtained by two different satellites allow us to use the
complementary properties of both satellites.

\section{Observations}
\label{sect:obs}
Results from the Low-Energy Concentrator Spectrometer
(LECS; 0.1--10 keV; Parmar et al.\ 1997), the Medium-Energy
Concentrator Spectrometer (MECS; 1.8--10 keV; Boella et al.\ 1997),
the High Pressure Gas Scintillation Proportional Counter (HPGSPC;
5--120 keV; Manzo et al.\ 1997) and the Phoswich Detection System
(PDS; 15--300 keV; Frontera et al.\ 1997) on-board \sax\ are
presented. All these instruments are coaligned and collectively
referred to as the Narrow Field Instruments, or NFI. The MECS consists
of two grazing incidence telescopes with imaging gas scintillation
proportional counters in their focal planes. The LECS uses an
identical concentrator system as the MECS, but utilized an ultra-thin
entrance window and a driftless configuration to extend the low-energy
response to 0.1 keV.

The non-imaging HPGSPC consists of a single unit with a collimator
that remained on-source during the entire observation. The non-imaging
PDS consists of four independent units arranged in pairs each having a
seperate collimator. Each collimator was alternatively rocked on- and
210\arcmin\ off-source every 96 s during the observation.

The region of sky containing Ser X-1 was observed by \sax\ on 1999
September 05 2:46 UTC to September 05 20:15 UTC, while GX 3+1 was
observed on 1999 August 30 18:33 UTC to August 31 11:41 UTC. Good data
were selected from intervals when the instrument configuration was
nominal, using the SAXDAS 2.0.0 data analysis package. LECS and MECS
data were extracted centered on the position of the sources using
radii of 8\arcmin\ and 4\arcmin, respectively. The exposures in the
LECS, MECS, HPGSPC, and PDS instruments are 8.9 ks, 30.8 ks, 30.3 ks,
and 14.0 ks for GX 3+1, while Ser X-1 was observed for 15.7 ks, 31.7
ks, 31.3 ks, and 15.4 ks, respectively.

Background subtraction for the imaging instruments was performed using
standard files, but is not critical for such bright
sources. Background subtraction for the HPGSPC was carried out using
data obtained when the instrument was looking at the dark Earth and
for the PDS using data obtained during intervals when the collimators
were offset from the source (see also Sect.\ \ref{results} for some
details regarding GX 3+1).

RXTE observation of Ser X-1 were obtained between 1999 September 05
3:33 UTC and September 05 14:06 UTC, while the observations of GX 3+1
were obtained on 1999 August 30 18:41 UTC to August 31 07:15 UTC. Only
data obtained with the Proportional Counter Array (PCA, Bradt et al.\
\cite{b:93}) were analysed. Most of the data were
obtained with 4 or 3 proportional counter units on, with time
resolutions ranging from 16 s (129 photon energy channels, effectively
covering 2--60 keV) down to 16$\mu$s using various timing modes
covering the 2--60 keV range.

\section{Results}
\label{results}

\begin{figure*}
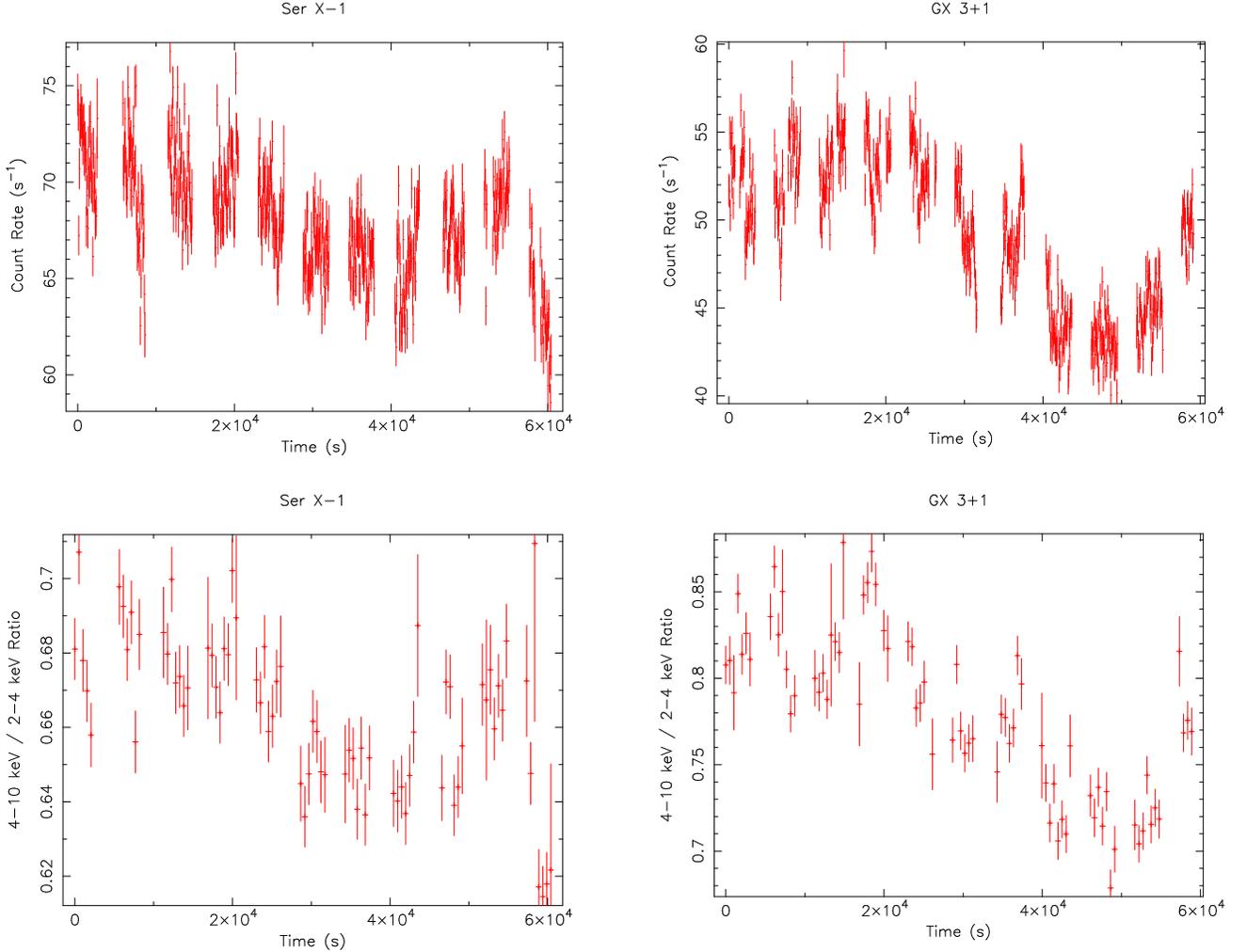

  \centerline{\hfill \mbox{\includegraphics[height=8cm,angle=-90]{H2350F1A.ps}}
\hspace{1cm}              \includegraphics[height=8cm,angle=-90]{H2350F1B.ps}}
\vspace{0.5cm}
\hfill \mbox{\includegraphics[height=8cm,angle=-90]{H2350F1C.ps} 
\hspace{1cm} \includegraphics[height=8cm,angle=-90]{H2350F1D.ps}}

  \caption[]{The Ser X-1 (left) and GX 3+1 (right) BeppoSAX MECS 2--10
             keV lightcurves using a binning time of 64s. At the
             bottom the hardness-ratios are plotted as a function of
             time in 512 s time bins} \label{fig:lightcurves}
\end{figure*}

The lightcurves of both sources were plotted and inspected: no bursts
are detected. The MECS count rate of Ser X-1 is $\sim$ 68 counts
s$^{-1}$, while the count rate of GX 3+1 varied between $\sim$42 and
$\sim$56 counts s$^{-1}$ (see Fig. \ref{fig:lightcurves}).

Using the default pipeline products the PDS spectrum of GX 3+1 shows
serious discrepancies (it was a factor 2--3 lower) with the HPGSPC
spectrum in the overlapping energy range. We found that the offset
spectra (which are used for the background spectrum) are obtained
while the PDS-detectors were observing a region of the sky containing
the two bright X-ray sources SLX\,1744-300 and SLX\,1744-299 in the
negative offset direction. We repeated the analysis using only the
positive offset direction and obtaining a spectrum which agrees much
better with the HPGSPC spectrum in the overlapping energy
range. However, the agreement is still not perfect; this is explained
by the fact that there are quite a few fainter sources in, or close
to, the field of view of the PDS in the positive offset-position. This
is unsurprising since GX 3+1 is very close to the galactic
center. Additionally, it might be expected that the gradient in the
galactic ridge emission might be important in the background
subtraction.  This means that the high-energy background subtraction
of GX 3+1 is rather uncertain.  For this reason we have excluded the
PDS spectrum of GX 3+1 for most of the analysis mentioned below.  We
note that, since the spectrum is soft, not much information is lost by
excluding the PDS spectrum from the analysis.  There are no such
problems for the Ser X-1 PDS spectrum (which is unsurprising since Ser
X-1 is located at a higher galactic latitude, no known bright sources
are present in the offset positions of the PDS).

Additionally the galactic ridge emission might contribute to the
HPGSPC spectrum, since dark-earth data is used for the subtraction.
We have taken the spectral shape and normalization for the galactic
ridge emission from Valinia \& Marshall (\cite{vm:98}) and find that
for GX3+1 the contribution from the galactic ridge emission is 1\% at
10 keV and 3\% at 20 keV (the upper boundary of the HPGSPC energy
range used), while the statistical uncertainty is $\sim$6\% at 20
keV. We conclude that, because of the brightness of the source, this
galactic ridge emission is not a problem in the background
subtraction. For Ser X-1 the situation is even better, since the
source is somewhat brighter, and more importantly, located at a higher
galactic lattitude.

We have checked for the presence of source in the FOV of the HPGSPC
and PDS and found that no sources brighter than 0.4\% of the intensity
of the sources of interest are present. Also no sources affecting the
analysis are present in the data of the imaging (LECS/MECS)
instruments. 

\subsection{RXTE results}
\subsubsection{GX 3+1}
RXTE data on GX 3+1 have been obtained starting at
roughly the same time and ending about 4 hours earlier than the \sax\
data (see Sect. \ref{sect:obs}).

\begin{figure}
  \centerline{\includegraphics[height=8cm,angle=-0]{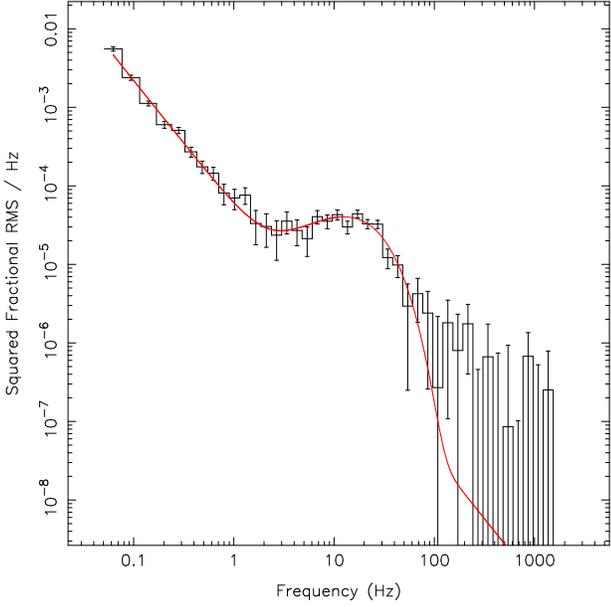}}
  \caption[]{The power spectrum obtained from the RXTE data for GX
3+1. The drawn line represents the best-fit ``Atoll'' model. Note the
  presence of the ``bump'' (HFN) around 20 Hz, typical of
  Atoll-sources in the banana state}
  \label{fig:power1}
\end{figure}

We have obtained the power spectrum by making FFTs using the event
data (E\_16us\_64M\_0\_1s) over the entire energy range with an
original time resolution of 16$\mu$s. We used 131072 points per FFT,
resampling the event data at a time resolution of 1/4096 s
(corresponding to a $\nu_{\rm Nyquist}$ of 2048 Hz), which results in
one power spectrum every 32 s.  We checked the power spectra for
strong variations as a function of time. We found none and analyzed
the average power spectrum. We find that this spectrum can be
adequately described with the ``Atoll'' model, which consist of very
low-frequency noise (VLFN; described by a power-law) and
high-frequency noise (HFN, described by a cutoff-powerlaw, see e.g.\
van der Klis \cite{vdk:95}). The obtained fit to the average power
spectrum results in a rms VLFN-strength of 7.5\% (0.01--1 Hz)
described by a power-law index, $\alpha$, of $\sim$ 1.6, HFN rms
(1--100 Hz) 3.5\% ($\alpha \sim$ -1.20, $\nu_{\rm cutoff} \sim 10.8$
Hz), see Fig.\ \ref{fig:power1}.  Those values (especially the
presence of the VLFN) are typical of Atoll sources in the lower banana
state (e.g. Hasinger \& van der Klis \cite{hk:89}). We have not
detected kilohertz QPO with an upper limit (at 95\%) to the
variability of $\sim$2.6\% rms.

\begin{figure}
  \centerline{\includegraphics[height=8cm,angle=-0]{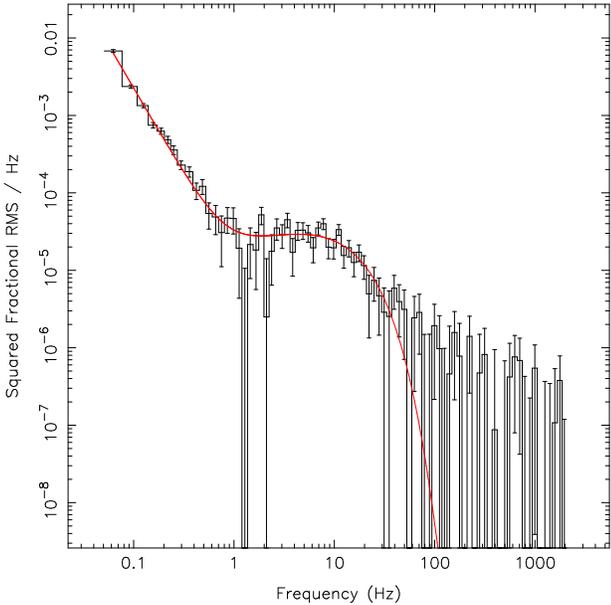}}
  \caption[]{The power spectrum obtained from the RXTE data for Ser
  X-1.  The drawn line represents the best-fit ``Atoll'' model. Note the
  presence of the ``bump'' (HFN) around 20 Hz, typical of
  Atoll-sources in the banana state}
  \label{fig:power2}
\end{figure}

Using the Standard 2f data we made a colour-colour diagram using time
bins of 16 s. This colour-colour diagram resembles the banana state
(high inferred mass accretion rate) of an ``Atoll'' source (which is
consistent with the power spectrum), since significant variations
occur in the soft colour (positively correlated with hard colour
variations). We therefore conclude that GX 3+1 was in a lower banana
state during the time of the RXTE and \sax\ observations.


We have extracted RXTE spectra and fitted these with the same {\sc
diskBB + compTT} model as used for the \sax\ data. Only Layer 1 data
were analyzed (since the source is relatively soft) for PCU 0,1, and
2. The response matrix was generated for this choice of detectors and
layers using {\sc ftools} 5.0. All obtained fit parameters are
consistent with the \sax -derived parameters within their uncertainties.

\subsubsection{Ser X-1}

RXTE data on Ser X-1 have been obtained starting $\sim$1 hour after
the start of the \sax\ observations and ending 6 hours before the end
of the \sax\ observations (see Sect. \ref{sect:obs}).

For Ser X-1 we used a similar procedure with event data obtained in
mode E\_125us\_64M\_0\_1s, but using the same resampling and length of
the FFTs.  We find that the power spectrum (Fig. \ref{fig:power2}) can
be well fitted using the same ``Atoll''-model, while the obtained
parameters are: VLFN rms of 23\% (0.01--1 Hz), $\alpha \sim$ 2.2, HFN
rms (1--100 Hz) 2.3\% ($\alpha \sim $ -0.5, $\nu_{\rm cutoff} \sim
9.3$ Hz).  The colour-colour diagram indicates that the source was in
the banana state (high inferred mass accretion rate), probably closer to
the upper banana state, since the HFN-strength is relatively low
(see e.g.\ van der Klis \cite{vdk:95}). No kilohertz QPO were detected
with an upper limit to the rms variability of $\sim$2.0\%

\begin{figure}
  \centerline{\includegraphics[width=8cm,angle=-0]{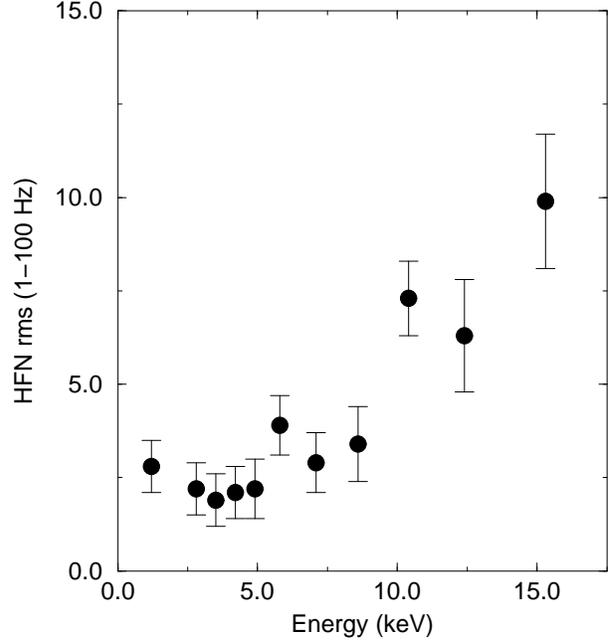}}
  \caption[]{The rms variability of the HFN component as a function
  energy for Ser X-1. The indicated energy is the central energy of
  the energy bins used for the power spectrum} \label{fig:hfn}
\end{figure}

We have also determined the energy dependence of the HFN strength as a
function of energy for Ser X-1. We find that the HFN rms increases
from $\sim$2\% at 3 keV to $\sim$9\% at 15 keV (see
Fig. \ref{fig:hfn}). We note that the increase in HFN noise strength
is roughly consistent with the larger contribution of the Comptonized
component to the total spectrum (i.e.\ the HFN noise gets diluted at
lower energies by the presence of the soft component). Due to the
lower count rate (and thus the poorer statistics) a similar plot for
GX 3+1 looks noisier; however, it is consistent with the plot obtained
for Ser X-1.

We have extracted RXTE spectra and fitted these with the same {\sc
diskBB + compTT} model as used for the \sax\ data. Only Layer 1 data
were analyzed (since the source is relatively soft). The choice of PCUs
and the method of generation of the response matrix was the same as
for GX 3+1.
The best-fit RXTE
values are consistent with the values (and uncertainties) determined
from the \sax\ data. The only parameter in which an appreciable
difference was found was the seed temperature of the electrons
($T_{0}$). The best-fit nominal value from the RXTE data was 0.32 keV,
but this parameter is poorly constrained (due to the lack of
low-energy response of the PCA) and is
therefore consistent with the value obtained from the \sax\ spectra.

\subsection{\sax\ spectral analysis}
\label{subsect:spectrum}

The overall spectra of GX 3+1 and Ser X-1 were first investigated by
simultaneously fitting data from all the \sax\ NFI.  The LECS and MECS
spectra were rebinned to oversample the full width half maximum of the
energy resolution by a factor 3 and to have additionally a minimum of
20 counts per bin to allow use of the $\chi^2$ statistic.  LECS
response matrices appropriate for the position and count rates of the
sources were generated and used.  The HPGSPC and PDS (for Ser X-1
only) spectra were rebinned using the standard techniques in SAXDAS.
Data were selected in the energy ranges 0.5--10.0~keV (LECS),
1.8--10~keV (MECS), 8.0--20~keV (HPGSPC), and 15--100~keV (PDS) where
the instrument responses are well determined and sufficient counts
obtained. Due to the high \nh\ we have not used data below 0.5 keV,
since this does not add a significant number of source counts. The
photoelectric absorption cross sections of Morisson \& McCammon
(\cite{mm:83}) and the solar abundances of Anders \& Grevesse
(\cite{a:89}) are used throughout.

\begin{table}
\caption[]{A comparison of models when fit to Ser X-1 and GX
             3+1. Indicated are the \rchisq\ values. pow: powerlaw;
diskBB: disk blackbody model (Mitsuda et al.\ \cite{m:84}); Gauss:
Gaussian; highe: high energy cutoff of the form: $\exp ((E_{\rm cutoff}
- E))/E_{\rm fold})$; cutoff: cutoff power law model ($E^{-\alpha}
\exp (-E/E_{\rm cutoff})$; compTT: Comptonizing model (Titarchuk
\cite{t:94}); BB: blackbody model; compST: Comptonizing spectrum
(Sunyaev \& Titarchuk \cite{st:80}); pexrav: exponentially cutoff
power law model reflected from neutral material (Magdziarz \&
Zdziarski \cite{mz:95})}
\begin{flushleft}
\begin{tabular}{lcc}
\hline\noalign{\smallskip}
Model              & Ser X-1 & GX 3+1 \\
\noalign{\smallskip\hrule\smallskip}
pow    + diskBB + Gauss    & 11.9 & 1.85 \\
pow*highe + diskBB + Gauss & 2.35 & 1.50 \\
cutoff + diskBB + Gauss    & 1.57 & 1.04 \\
compTT + diskBB + Gauss    & 1.54 & 0.99 \\
compTT + diskBB            & 3.52 & 1.88 \\
compTT +   BB   + Gauss    & 1.51 & 0.99 \\
compST + diskBB + Gauss    & 1.73 & 1.23 \\
compST +   BB   + Gauss    & 1.55 & 1.12 \\
pexrav + diskBB + Gauss    & 1.49 & 1.47 \\
\noalign{\smallskip\hrule\smallskip}
\end{tabular}
\end{flushleft}
\label{tab:models}
\end{table}


Initially, simple models were tried, including absorbed power-law,
and cutoff power-law models, but these models gave a poor description
to the obtained data.
Factors were included in the spectral fitting to allow for normalization 
uncertainties between the instruments. These factors were constrained
to be within their usual ranges during the fitting.

\begin{figure*}
  \centerline{\mbox{\includegraphics[height=8cm,angle=-90]{H2350F5A.ps}}
\hspace{1cm}              \includegraphics[height=8cm,angle=-90]{H2350F5B.ps}}
  \caption[]{The overall Ser X-1 (left) and GX 3+1 (right) \sax\ NFI spectra together with the best-fit 
             {\sc compTT} and disk-blackbody plus Fe Line model fits
             (see Table~\ref{tab:spec_params}). In the lower panels
the contribution to the \chisq\ is plotted}
  \label{fig:spectrum}
\end{figure*}

\begin{figure}
  \centerline{\includegraphics[height=8cm,angle=-90]{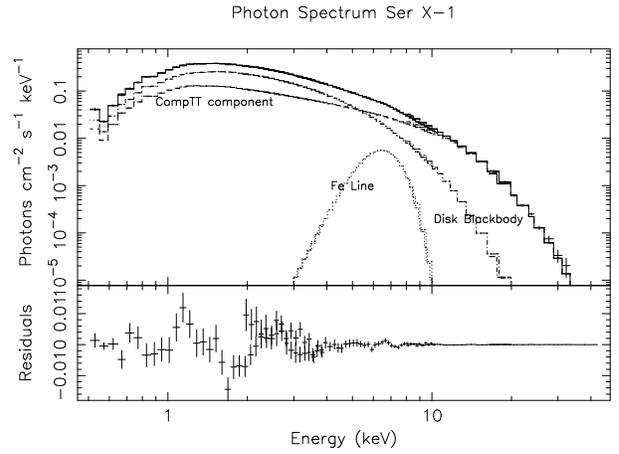}}
  \caption{The photon spectrum for Ser X-1, illustrating the
  broad-band spectral shape. Since the spectral parameters for GX
  3+1 are rather similar to those of Ser X-1, the photon spectrum is 
  comparable. The separate components (disk blackbody,
  comptonized spectrum and iron line) are indicated}
  \label{fig:spectrum_unfolded}
\end{figure}

\begin{table}
\caption[]{Fit results for Ser X-1 and GX 3+1. $T_{0}$ is the input
seed photon temperature for the Comptonization model; $\tau$ the
optical depth of the Comptonizing plasma, which has a temperature $T$.
$ kT_{\rm in}$ and $R_{\rm in} \sqrt{\cos \Theta}$ are the parameters 
describing the disk blackbody model: the temperature at the inner disk
and the ``inner disk radius'' (i.e.\ the normalization).}
\begin{flushleft}
\begin{tabular}{lcc}
\hline\noalign{\smallskip}
Parameter & Ser X-1 & GX 3+1 \\
\noalign{\smallskip\hrule\smallskip}
${\rm N_H}$ 10$^{2}$ \hcm  & 0.50$\pm ^{0.02} _{0.03}$  & 1.59$\pm ^{0.07}_{0.12}$ \\
$T_{0} $ (keV) & 0.15$\pm ^{0.06}_{0.15}$ & 0.49$\pm ^{0.05} _{0.03}$ \\
$T     $ (keV) & 2.52$\pm$ 0.07 & 2.71$\pm ^{0.10} _{0.27}$ \\
$ \tau $ & 9.7$\pm ^{2.0} _{1.0} $  &  6.1$\pm ^{1.5} _{3.8}$ \\
$ kT_{\rm in}$ (keV) & 1.46$\pm$ 0.05 & 1.95$\pm ^{1.02} _{0.50}$ \\
$ R_{\rm in} \sqrt{\cos \Theta}$ (km) &  6.8$\pm 0.8 $ & 2.8$\pm ^{1.4} _{1.2}$ \\
Fe E (keV) & 6.46$\pm ^{0.12} _{0.14}$  & 6.1$\pm ^{0.40} _{0.45}$ \\
Fe $\sigma$ (keV) & 0.98$\pm ^{0.15} _{0.13}$ & 1.00$\pm ^{0.35} _{0.28}$ \\
Fe EW (eV) & 275$\pm ^{75} _{55}$  & 200$\pm ^{200} _{85}$ \\
$\chi ^2$/dof & 152.8/99 & 88.1/89 \\
Flux (2--10 keV)  & 5.4  & 4.7 \\
Flux (10--50 keV) & 1.2  & 0.9 \\
Flux (1--20 keV)  & 7.9  & 5.0 \\
Flux (20--200 keV)& 0.08 & 0.06\\
F$_{\rm DBB}$/F$_{\rm total}$ (0.1--10 keV) & 0.54 & 0.40 \\
\noalign{\smallskip\hrule\smallskip}
\multicolumn{3}{l}{\footnotesize Assumed distance to GX 3+1 is 4.5 kpc
(Kuulkers \& van}\\
\multicolumn{3}{l}{\footnotesize der Klis 2000).}\\
\multicolumn{3}{l}{\footnotesize $ R_{\rm in} \sqrt{\cos \Theta}$ for
Ser X-1 is expressed in D$_{\rm 10 kpc}$}\\
\multicolumn{3}{l}{\footnotesize Flux is in units of 10$^{-9}$ erg
s$^{-1}$ cm$^{-2}$ and unabsorbed}\\
\multicolumn{3}{l}{\footnotesize All uncertainties have been obtained by
$\Delta \chi^{2}=2.71$}\\
\end{tabular}
\end{flushleft}
\label{tab:spec_params}
\end{table}

A model consisting of a thermal component (dominating the bulk of the
emission below 1 keV) and a thermal Comptonization ``tail'' provides
an acceptable description of the broadband spectrum in both sources
(as typically observed in bursters; Guainazzi et al. 1998; in't Zand
et al. 1999; Barret et al. 2000). Positive residuals around
$\sim$6.5~keV are suggestive of the presence of fluorescent iron
emission line, which is detected for the first time in these
sources. The thermal component can be described either by a single
temperature blackbody or by a multi-temperature disk blackbody.

The difference in \rchisq\ between disk-blackbody and blackbody models
is rather small. The models do not provide a statistically acceptable
\chisq . However, after adding a 2\% systematical uncertainty to the
LECS and MECS data, the \rchisq\ is much closer to one. Therefore the
question whether the best-fitting models give a good statistical
description to the data is hard to answer, but we can say that the 2\%
systematical uncertainties are roughly in-line with the calibration
uncertainties for bright sources of the \sax\ 
instruments. Additionally, some spectrals changes might be expected as
a function of intensity, which might contribute to the high \chisq
(however, the intensity selected spectra of GX 3+1 do not show strong
evidence for this effect).  All quoted \chisq\ values and
uncertainties in the text and Tables refer to values obtained with the
2\% systematic uncertainty included.  Most of the contribution to the
high \chisq\ is coming from the 4-7 keV region in the MECS, and looks
highly structured. We note that this region includes the instrumental
Xenon L-edge and the Fe line, which might be poorly modeled. We
believe that the structured shape of the residuals argues against the
possibility that the continuum has been modeled incorrectly (e.g., due
to the presence of a reflected component), since this would tend to
give more smoothly varying residuals.

In order to compare the spectra of the sources we have used for the
detailed fitting the models which gives, on average, the best fit to
both spectra. This model is the {\sc compTT + diskBB + Gaussian}
model. The choice between {\sc blackbody} and {\sc diskBB} is
arbitrary. Additionally the obtained values for the inner disk radius
($ R_{\rm in} \sqrt{\cos \Theta}$) are rather small. However,
following Merloni et al.\ (\cite{m:00}), we believe that the obtained
value is not a reliable estimate (but an underestimate) of the true
inner radius of the disk. Similar results were obtained by Barret et
al.\ (\cite{b:00}) for KS\,1731--260.

Note that the best-fit model to the Ser X-1 spectrum contains a {\sc
pexrav} model component (see Table \ref{tab:models} and Sect.\
\ref{sect:discussion}). However, the difference in \chisq with respect
to the model consisting of a disk blackbody and {\sc compTT} component
is not statistically significant. The obtained value for the
``relative reflection'' is 1.28$\pm$0.03 (with $\cos$ inclination
fixed to the default value of 0.45), which implies a rather high
contribution of the reflected component.

Since the variations in intensity and colours are more pronounced for
GX 3+1 than for Ser X-1, we have attempted to study the spectra
obtained at high and low intensities. We have obtained a spectrum at
``low'' count rates ($<$49 cts s$^{-1}$ in the MECS) and ``high''
count rates ($>$49 cts s$^{-1}$), using time bins of 256 s. There
seems to be a small, but significant, difference between these spectra
of GX 3+1 obtained at low and high count rates (within the \sax\
observation), which is also reflected in the colours.  However, the
spectral fitting results in parameters which are, within their
uncertainties, the same between the two spectra.  The spectral
differences between the two spectra are very small, and at the limit
which can be resolved by \sax .  Additonally no large change in the
observed \rchisq\ was obtained. This justifies using the average
spectrum for the purpose of determining the broad-band shape of the
continuum.  Therefore we have not attempted a similar approach to Ser
X-1 (where the differences in colours and count rates are even less
than in GX 3+1).

The observed Equivalent Widths (EW) of the Fe line is $\sim$200 eV for
GX 3+1. Especially in the case of GX 3+1 the energy and the width of
the iron line is not well constrained. We note that that a broad Fe
line gives a significantly better fit to the spectra for both Ser X-1
and GX 3+1: for Ser X-1 the \chisq /degrees of freedom (dof) decreases
from 238.7/100 (for a zero-width line) to 152.9/99 (for a broad line),
while for GX 3+1 the change is from are 114.2/90 to 88.1/89. The
decreases in \chisq\ are highly significant.  It is rather likely that
the shape of the Fe line is more complex than a simple Gaussian
(i.e. blend of different energies, presence of an edge, or broadening
by Comptonization). The fits tend to favour a broad line (see Table
\ref{tab:spec_params}), which might be caused by
Comptonization. However, this possible complexity is not well
constrained by our data.


\section{Discussion}
\label{sect:discussion}

We have observed GX 3+1 and Ser X-1 in the 0.1--30~keV energy range
using \sax.  We find that the spectra are best described by a
combination of a {\sc compTT} model and blackbody-like model (either a
blackbody or a multi-colour blackbody model {\sc diskBB}). From the
RXTE timing and colour-colour diagram we infer that GX 3+1 is in the
lower banana state, implying a $\dot M$ of around 0.15 L$_{\rm Edd}$,
while Ser X-1 is in the upper banana state, implying a slightly higher
mass accretion rate. The absence of kilohertz QPO is consistent with
the fact that kilohertz QPO are not normally seen in the banana-type
state in other sources (for a list of similar upper limits see van
Straaten et al.\ 2000).

Comparison of the spectra of the two sources discussed here with the
spectra of various low-mass X-ray binaries (see Barret et al.\
\cite{b:00}), reveals a striking resemblance in terms of
fit-parameters with KS1731-260. This is not suprising, since this
source was also in the banana state during the RXTE observations. The
obtained fit-parameters are very similar.  Comparing our two sources
with another well-studied atoll-source like 4U\,1820--30 we see that
the spectral parameters are compatible when 4U\,1820--30 is rather
bright (e.g.\ the observation reported in Piraino et
al. \cite{p:98}). Interestingly 4U\,1820--30 is not detected at high
($\sim$100 keV) X-ray energies (see e.g.\ Bloser et al.\
\cite{bloser:00}), even at its lowest accretion rates.

The observations presented here provide additional evidence for a
strong relation between the luminosity and the presence of a hard tail
which can be modeled by thermal Comptonization with an electron
temperature ranging from $\sim$30 keV at the lowest accretion rates to
$\sim$ 2.5 keV at higher accretion rates ($\sim$0.15 L$_{\rm
Edd}$). Thus the broad-band spectral properties of X-ray burst sources
seem to be governed, to first order, by the mass accretion rate and
appear to be very similar between sources.


The observed optical depth ($\tau$), electron temperature and the
diskblackbody spectral parameters are rather similar to X1820-303, the
brightest source in the sample of globular cluster sources as
discussed in Guainazzi et al.\ (2000). This is consistent with the
observed relation between luminosity and e.g. $\tau$ (Guainazzi et
al. 2000) and the fact that these sources are expected to accrete at a
$\sim$10--30\% of the Eddington limit. Using the correlation between
the observed $\tau$ and accretion rate expressed in Eddington
luminosity (Fig.\ 4 in Guainazzi et al. 2000) we can estimate the
Eddington luminosity of Ser X-1 and hence its distance. The observed
$\tau$ would put it a luminosity of $\sim$0.15 L$_{\rm Edd}$, which
would place Ser X-1 at a distance of $\sim$6 kpc.  A similar exercise,
using the observed $\tau$, would place GX 3+1 at a distance of $\sim$5
kpc, in remarkable agreement with the distance of 4.5 kpc determined
by Kuulkers \& van der Klis (2000) using a unique radius-expansion
burst from this source. However, this remarkable agreement may be
fortuitous, since a more complete sample of globular cluster
sources (Sidoli et al.\ \cite{s:00}) shows a less clear correlation
between luminosity and $\tau$ than in Guainazzi et al.\ (2000).

Alternatively the observed $\tau$ from the spectrum of GX 3+1 can be
used to estimate the luminosity in terms of L$_{\rm Edd}$, using the
correlation in Guainazzi et al. (2000). We found a luminosity of
$\sim$0.1 L$_{\rm Edd}$. This value, when the source is in the lower
banana state is in line with current thoughts about the mass-accretion
rate in this state (van der Klis \cite{vdk:95}).  For Ser X-1 we infer
a slightly higher $\dot M$, since it probably is higher ``up'' the
banana-curve, which is consistent with the inferred luminosity of
$\sim$0.15 L$_{\rm Edd}$ from $\tau$.

The energy dependence of the rms variability of the high-frequency
noise (Fig.\ \ref{fig:hfn}) shows that the HFN is ``hard'', i.e.\ the
variability increases towards higher energies. This has been observed
before in e.g.\ 1E1742.5--3045 (Olive et al. \cite{o:98}, in the
island state), GX 3+1 (Makishima et al.\ \cite{m:89}), and 4U 1820--30
(Dotani et al.\ \cite{d:89}). The energy dependence is roughly
consistent with a $\sim$8\% variability in the Comptonized component,
which, at lower energies, progressively gets diluted by the soft
component (the (disk-)blackbody). However the spectral decomposition
is not unique and the statistical quality of the rms variability as a
function of energy is not good enough to associate the variability
uniquely with one spectral component. e.g.\ it is conceivable that (an
energy-dependent) part of the disk-blackbody component also
contributes to the variability.

It is interesting to interpret the spectrum in terms of the boundary
layer model (e.g.\ Popham \& Sunyaev \cite{ps:00}; earlier work by
e.g.\ Kluzniak \& Wilson \cite{k:91}). In this model the 
dominant source of high-energy photons is expected to be the boundary
layer where the accretion disk meets the (weakly magnetized) neutron
star. Roughly speaking this model predicts a Comptonized component
together with a thermal component. Our results (and those of e.g.\ 
Barret et al.\ \cite{b:00}) are in global agreement with the
predictions of this model. Also the fact that the observed variability
of the high-frequency noise is consistent with most of the variability
originating in the Comptonized component, argues in favour of the
model.

The observed Fe line energy is consistent with neutral or moderately
ionized iron.  However, care should be taken when interpreting the
energy of the Fe line, since the addition of an edge significantly
shifts the energy (to a value consistent with higly ionized Fe i.e.\
6.7 keV). Following the discussion in Barret et al.\ (\cite{b:00}) the
most likely location of the material emitting the Fe line is in the
accretion disk.  A strong Fe line should then be accompanied by a
Compton reflected spectrum. It is indeed the case that a fit with a
reflected component gives a better fit (although it is not
significantly better than the disk blackbody + compTT fit) to the Ser
X-1 spectrum (where the Fe line is strong, $\sim$275 eV).  However,
the quality of our data does not allow us to draw any definitive
conclusions about the presence or absence of reflection.  The obtained
width ($\sigma$) of the Fe line is significantly higher than 0. This
might be caused by Comptonization of the Fe line photons, or by other
effects which cause broadening of the Fe line.

Summarizing we have observed two burst sources (GX 3+1 and Ser X-1)
with similar spectral parameters (temperature of Comptonizing plasma
$\sim$2.5 keV with a large (6--10) optical depth). This is in contrast
with low state spectra of burst sources which have a much higher
temperature ($\sim$30 keV) and a smaller optical depth.

\begin{acknowledgements}
The \sax\ satellite is a joint Italian-Dutch programme. 
We thank the \sax\ and RXTE planners (Donatella Ricci and Evan Smith)
for making the simultaneous observations possible. We thank Arvind
Parmar for a careful reading of the manuscript. Jean Swank is
thanked for making the RXTE observations possible. We thank the
referee, Phil Kaaret, for very useful comments.

\end{acknowledgements}

\end{document}